# A Strategy for Improving the Interfacial Crystallinity and Carrier Mobility of SnO$_2$ Porous Nanosolids

Chunhong Luan[1,2], Chao wang[1], Bao Xu[1], Yujing Geng[2], Liangmin Zhang[3], Qilong Wang[4], Deliang Cui[2]*

1 School of Microelectronics and Solid-State Electronics, University of Electric Science and Technology of China, Chengdu 611731, P.R. China

2 State Key Lab of Crystal Materials, Shandong University, Jinan 250100, P.R. China

3 Arkansas Center for Laser Applications and Science, Department of Chemistry and Physics, Arkansas State University, Arkansas 72467, USA

4 Key Laboratory for Special Functional Aggregated Materials of Education Ministry, School of Chemistry & Chemical Engineering, Shandong University, Jinan 250100, P.R. China

**Abstract** SnO$_2$ porous nanosolid (PNS) was prepared by a solvothermal hot-press method, and a new strategy was developed to improve its interfacial crystallinity and carrier mobility. It was found that the carrier mobility of SnO$_2$ PNS was improved after being calcined at 500 °C in high-pressure oxygen. Furthermore, the mobility was greatly increased by calcining SnO$_2$ PNS at 350 °C for 12 h in high-pressure oxygen, and the highest mobility reached 35 cm$^2$/Vs. On the other hand, the complex impedance spectra of the samples revealed that the annihilation of oxygen vacancies mainly happens within the interfacial region among SnO$_2$ nanoparticles during the calcinations process in high-pressure oxygen. As a result, the interfacial crystallinity was improved and carrier mobility was increased. Based on the analysis of experimental data, a simple model was proposed to explain the above phenomena. And the simple model may find applications in improving the carrier mobility of other oxide semiconductors used in photovoltaic cells.

## 1. Introduction

Up to now, Grätzel solar cell [1], which is constructed by using TiO$_2$ nanoparticles as the electron acceptors, possesses the highest photoelectric conversion efficiency. However, there are high density structural defects on the surface of TiO$_2$ nanoparticles, resulting in the formation of localized state for the charge carriers. These structural defects usually serve as the electron trap centers, decreasing their mobility and resulting in the recombination of electrons and holes. As a result, dark current of the solar cells obviously increases and the photoelectric conversion efficiency decreases accordingly. In order to further improve the performance of solar cells, intensive efforts have been devoted to improving the crystallinity of TiO$_2$ nanoparticles and searching for alternative oxide semiconductors.

Among the oxides used in solar cells, tin dioxide (SnO$_2$) is a wide band gap ($E_g$=3.6 eV) n-type semiconductor. Because of its excellent properties, SnO$_2$ has been widely used in dye-sensitized solar cells [2-7], transparent conducting films [8-10], sensor devices [11-13], field effect transistors etc [14]. When being used in solar cells, SnO$_2$ should have high crystallinity and electron mobility, so that better photoelectric conversion performance can be achieved. Fortunately, the mobility of SnO$_2$ is much higher than that of TiO$_2$ [14-16], thus it is a promising candidate for fabricating high-performance dye-sensitized solar cells. However, due to the fact that SnO$_2$ used in solar cells is usually in the form of nanoparticles, the interfacial crystallinity among SnO$_2$ nanoparticles plays an important role in determining the carrier mobility [17]. This result reveals that developing new route to improve the crystalline perfection of interfacial region is very important.

Herein, we find a rule that the carrier of the SnO$_2$ porous nanosolid (PNS) is related to the calcining atmosphere. As a result, we report a strategy of calcining SnO$_2$ porous nanosolid (PNS) in high-pressure oxygen, by which both the interfacial crystallinity and carrier mobility of SnO$_2$ PNS are greatly improved. It is believed that this route may find applications in improving other oxide semiconductors containing oxygen vacancy defect (for example, TiO$_2$, ZnO, etc), which are also used in photovoltaic cells.

## 2 Experimental

2.1 Materials

The starting materials used in the experiments were SnO$_2$ nanoparticles (with particle size of 50~70 nm, purchased from Luoyang Institute of Materials, Henan, China) and dioxane (C$_4$H$_8$O$_2$, A.R.).

2.2 Preparation of SnO$_2$ porous nanosolid

SnO$_2$ PNS was prepared by using a solvothermal hot-press (SHP) autoclave (Fig. 1) [18].

Firstly, 3 g SnO$_2$ nanoparticles were mixed with 5 ml dioxane and ground for 3 h in a planetary ball mill at a rate of 180 rpm. Then the resultant mixture was mounted into the SHP autoclave. When being heated to 100 ºC at a rate of 2.5 ºC /min, a constant pressure of 60 MPa was applied by the pistons of the autoclave. Afterward, the temperature was further increased at the same rate to 200 ºC and kept constant for 3h. After the autoclave was cooled to room temperature, the pressure was released and a green SnO$_2$ PNS was obtained. By calcining the green SnO$_2$ PNS at 500 ºC in air for 2 h, a well-defined SnO$_2$ PNS was obtained.

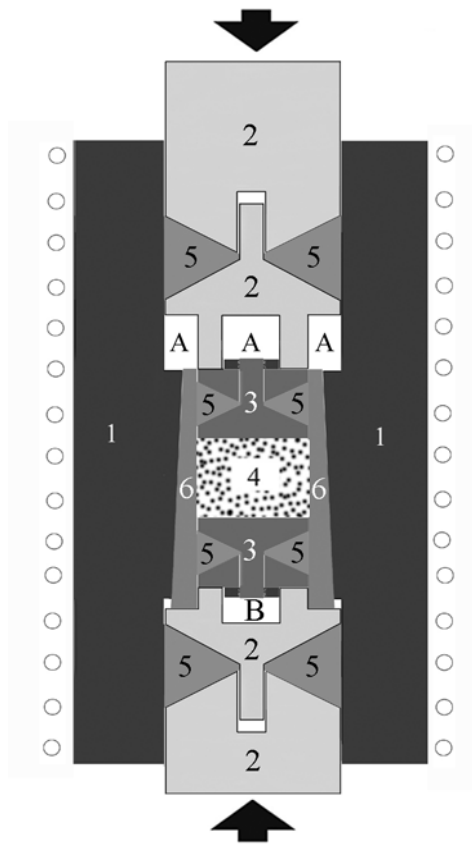

Fig. 1. Schematic diagram of a solvothermal hot-press autoclave.

(1) cylinder, (2) pistons, (3) gaskets, (4) sample, (5) teflon rings, (6) internal cylinder. A and B are the spaces for storing pore-forming solvent

2.3 Characterization of the samples

SnO$_2$ PNS is an intermediate state between nanoparticles and dense nanoceramics, which possesses both high reactivity of nanoparticles and strength of nanoceramics. The SnO$_2$ PNS contains interconnected SnO$_2$ nanoparticles, which form a framework with numerous pores (Fig. 2). Pore size distribution of the SnO$_2$ PNS was examined on a Quantachrome Pore Master-60 mercury intrusion porosimetry at 20 ºC. It is found that the porosity of SnO$_2$ PNS is 49.5% and its primary pore diameter is 50~100 nm (Fig. 3). The carrier concentration and mobility of SnO$_2$ PNS were determined by Hall measurement, which was carried out on an ET9000 series electrical transportation measuring system. Au electrodes were sputter-coated onto the corners of square-shaped samples (7.5×7.5×0.5 mm) in a four-point van der Pauw configuration [19]. Impedance spectroscopy measurements were carried out on an Agient 4294A & E499A impedance analyzers with overlapping ranges, and the measurement frequency range is 40 Hz~110 MHz. The pellets (with diameter of 23 mm and thickness of 2.5 mm) were sputter-coated with gold electrodes on the opposite surfaces, thereby forming parallel plate capacitor geometry.

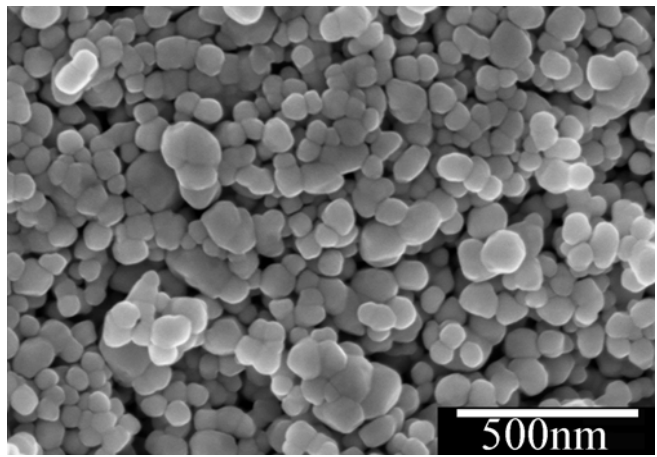

Fig. 2. SEM image of a SnO$_2$ PNS

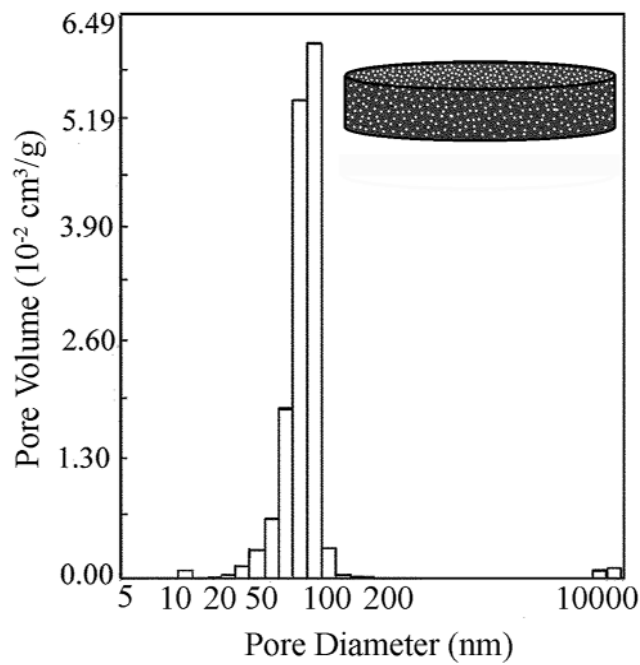

Fig.3 Pore size distribution chart of a SnO$_2$ PNS

Inset shows the schematic structure of a SnO$_2$ PNS, white dots denote the pores

**3 Results and Discussion**

3.1 Modulating carrier mobility of SnO$_2$ PNS by calcining in different atmospheres

SnO$_2$ possesses n-type semiconducting characteristic, which is resulted from the oxygen vacancies in SnO$_2$ nanoparticles. Generally, both the conductivity and carrier mobility of a SnO$_2$ film/ceramics are mainly determined by the crystallinity of interfacial region, because of the large particle size and comparatively thin depletion layer [20]. Besides, the concentration and distribution of oxygen vacancies in SnO$_2$ nanoparticles can be modulated by thermal-treatment, and this is especially true within the surface layer of SnO$_2$ nanoparticles [21, 22]. So, both the concentration and mobility of the carriers can be modulated by calcining SnO$_2$ in different gases.

Generally, oxygen molecule is chemisorbed at the surface of SnO$_2$, and it transforms into $O_2^-{}_{ads}$ by capturing an electron from SnO$_2$ below 160 ºC. With the temperature further increasing above 160 ºC, another electron is captured and $O_2^-{}_{ads}$ transforms into $O^-{}_{ads}$. There is a strong tendency for $O^-{}_{ads}$ to grab another electron from SnO$_2$, transforming into $O^{2-}{}_{ads}$ and subsequently becoming lattice oxygen by overcoming Madelung potential [23].

On the basis of the above results, we investigated the effect of calcining in different atmospheres on the mobility of SnO$_2$ PNSs. In our experiments, the SnO$_2$ PNSs were calcined at 500 ºC for 2 h in air, nitrogen and oxygen, respectively, thus samples A1, A2 and A3 were prepared. As an example, Fig. 4 shows that the I-V curves of sample A1 possess quite good linearity, indicating the homogeneity of this sample and the ohmic contact between the electrodes and SnO$_2$ PNS.

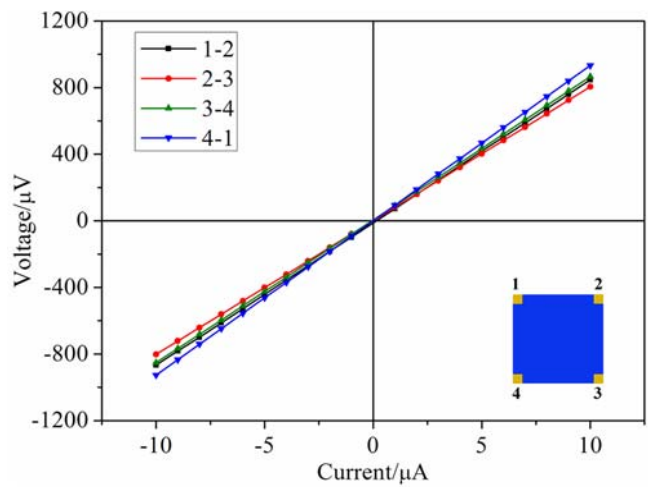

Fig. 4. I-V curves of sample A1

Table 1 shows the results of Hall measurement of samples A1, A2 and A3. Comparing to A1 and A3, sample A2 possesses higher carrier concentration $n$ and lower mobility $\mu_H$. In fact, the SHP process that is used to prepare $SnO_2$ PNS is completed in a sealed oxygen-deficient space (also see Fig. 1), thus oxygen vacancy defects will inevitably exist both at the surface of $SnO_2$ nanoparticles and in grain boundaries. Besides, there are large amount of defects in the starting $SnO_2$ nanoparticles as well. The oxygen vacancies play two important roles, the first role is to provide mobile electrons and making $SnO_2$ an n-type semiconductor, the second one is to decrease the carrier mobility. When being calcined at 500 ºC in $N_2$, the existing oxygen vacancies in $SnO_2$ cannot be effectively eliminated, contrarily, more oxygen vacancies can be formed in $SnO_2$ nanoparticles [24]. The results were also verified in our previous work [25]. As a result, the carrier concentration of sample A2 is quite high and its mobility is low. Compared to A1 and A2, appreciable amount of oxygen vacancies in A3 must have annihilated after being calcined in $O_2$, resulting in the increase of carrier mobility.

Table 1. Carrier concentration and mobility of $SnO_2$ PNSs calcined at 500 ˚C for 2 h in different atmospheres

| $SnO_2$ PNSs | Atmospheres | Resistivity (Ω·cm) | $n \times 10^{-17}$ (1/cm$^3$) | $\mu_H$ (cm$^2$/Vs) |
|---|---|---|---|---|
| A1 | Air | 8.52 | 1.29 | 5.67 |
| A2 | $N_2$ | 2.74 | 5.83 | 3.91 |
| A3 | $O_2$ | 4.50 | 2.30 | 5.98 |

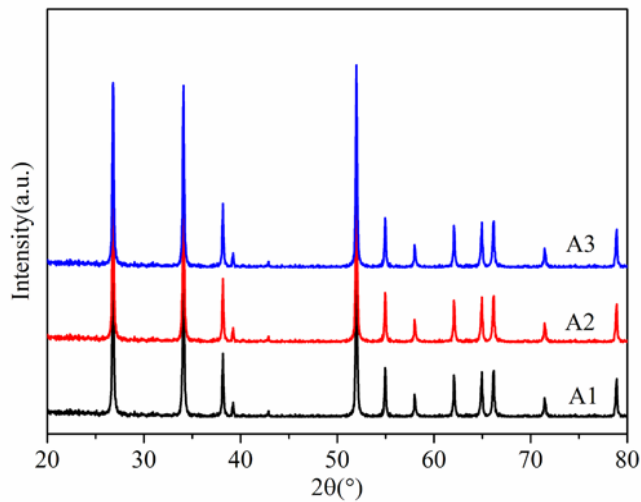

Fig. 5. XRD patterns of samples A1, A2 and A3

In order to understand the mechanism correlated to the above phenomenon, the XRD patterns are presented in Fig.5, and then, the complex impedance spectra of samples A1, A2 and A3 were collected. In Fig.5, all the peaks can be indexed to rutile $SnO_2$, and no obvious change has been observed after comparing these XRD patterns. This phenomenon suggests that both the phase structure and average particle size of $SnO_2$ nanoparticles remain unchanged after the calcining process. It means the calcining process might only change the properties of the surface and interface of the $SnO_2$ PNSs.

In the complex impedance measurement, all the $SnO_2$ nanoparticles, grains boundaries and the interface between electrode and $SnO_2$ PNS contribute to the impedance spectra [26]. Besides, the capacitance in the equivalent circuit is often substituted by a constant phase angle element (CPE), which represents the non-ideal capacitance characteristic. Therefore, we proposed an equivalent circuit (RC)(RQ)(RC) (Fig. 6, Q denotes the CPE) for the $SnO_2$ PNS and fit the measurement results using Zsimp Win 3.10 software.

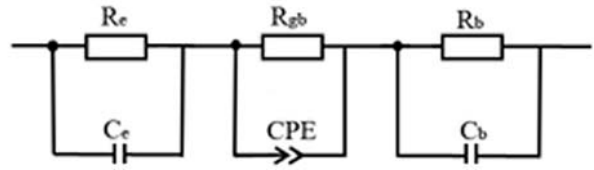

Fig. 6. The equivalent circuit of a $SnO_2$ PNS

The complex impedance spectra of samples A1, A2 and A3 are shown in Fig. 7, and the fitting results are listed in Table 2. The diameter of $SnO_2$ nanoparticles $D$ in our samples is 50~70 nm, while the Debye length $L_D$ of $SnO_2$ is about 3 nm, thus it can be regarded that $D>>L_D$, so the grain boundary plays a dominating role in the carrier transportation [27, 28]. In fact, the grain boundary resistance ($R_{gb}$) of $SnO_2$ PNS remarkably increased after being calcined in air and oxygen. Furthermore, Sample A3, prepared by calcining in oxygen, has the highest $R_{gb}$, revealing that more oxygen vacancies annihilated in this sample. This phenomenon will inevitably result in the decrease of carrier concentration and increase of depletion layer thickness within the interfacial region [29], which subsequently causes the increase of grain boundary resistance $R_{gb}$. On comparison, the $R_b$ values of these samples changes much less. This result should be resulted from the fact that the diffusion of oxygen within $SnO_2$ nanopartilcles is much slower than the oxygen exchange reaction on the surface [30].

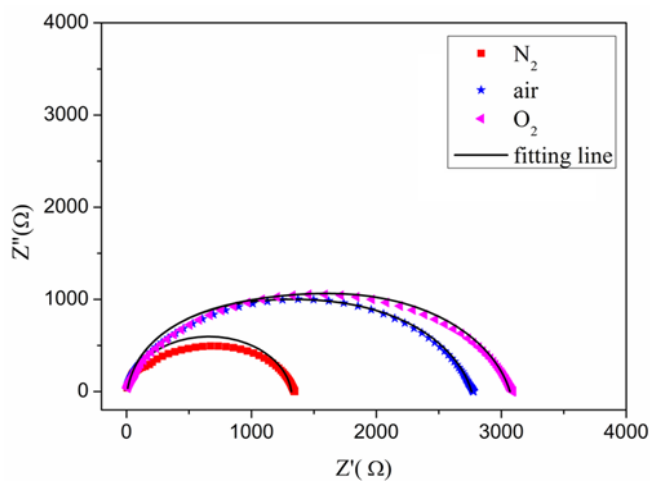

Fig. 7. Impedance spectra of samples A1, A2 and A3

Table 2. Fitting results of impedance spectra of the samples calcined at 500 ˚C for 2 h in different atmospheres

| $SnO_2$ PNSs | Atmospheres | $R_e$ (Ω) | $R_{gb}$ (Ω) | $R_b$ (Ω) |
| --- | --- | --- | --- | --- |
| A1 | air | 5.39 | 2430 | 1155 |
| A2 | $N_2$ | 9.25 | 1133 | 594.3 |
| A3 | $O_2$ | 6.79 | 2882 | 1102 |

3.2 Increasing the carrier mobility by calcining $SnO_2$ PNS in high-pressure oxygen

The above results reveal that structural defects within grain boundary in $SnO_2$ PNS can be restored by calcining in oxygen, thus the carrier concentration decreases and the corresponding mobility increases to some extent. Then what will happen to $SnO_2$ PNS if it were calcined in high-pressure oxygen? Here we conducted a series of experiments in which $SnO_2$ PNS was calcined at 500 ºC for 2 h under different oxygen pressures. Just as shown in Table 3, the carrier concentration decreases monotonously when the pressure of oxygen increasing from 0.1 MPa to 10 MPa, while the mobility increases from 5.98 to 9.08 $cm^2$/Vs (also see Fig. 8). This result can be attributed to the annihilation of oxygen vacancies and the improvement of crystallinity of the interfacial region among $SnO_2$ nanoparticles. In fact, accompanying the increase of oxygen pressure, more and more oxygen molecules are chemisorbed on the surface of $SnO_2$ nanoparticles while $SnO_2$ PNS is calcined in oxygen. These chemisorbed oxygen molecules will subsequently convert into lattice oxygen by grabbing electrons from $SnO_2$ nanoparticles, resulting in the improvement of their crystallinity and the annihilation of oxygen vacancies [31]. As a result, the carrier concentration is decreased and the mobility is improved.

Table 3. Carrier concentration and mobility of SnO$_2$ PNSs calcined at 500 ˚C for 2 h in high-pressure oxygen

| SnO$_2$ PNSs | O$_2$ pressure (MPa) | Resistivity (Ω·cm) | $n \times 10^{-17}$ (1/cm$^3$) | $\mu_H$ (cm$^2$/Vs) |
|---|---|---|---|---|
| P1 | 0.1 | 4.50 | 2.30 | 5.98 |
| P2 | 2.0 | 4.53 | 2.25 | 6.11 |
| P3 | 4.0 | 3.89 | 2.39 | 6.70 |
| P4 | 6.0 | 5.74 | 1.57 | 6.94 |
| P5 | 8.0 | 7.15 | 1.14 | 7.66 |
| P6 | 10.0 | 7.16 | 0.96 | 9.08 |

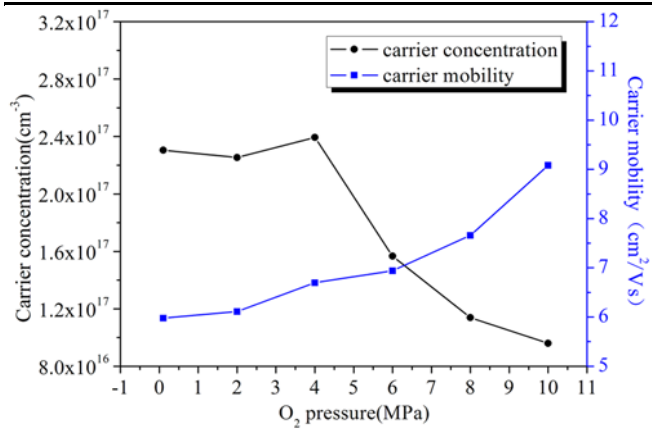

Fig. 8. Variation of carrier concentration and mobility of SnO$_2$ PNSs

The samples are calcined at 500 ˚C for 2 h in high-pressure oxygen.

Here the complex impedance spectra of SnO$_2$ PNSs are used again to analyze the mechanism of the above phenomenon. Fig. 9 presents the impedance spectra of SnO$_2$ PNSs calcined at different oxygen pressures, and the fitting results of these spectra are shown in Table 4. It can be clearly seen that R$_{gb}$ increases to a large extent with the oxygen pressure increasing from 0.1 MPa to 10 MPa, while R$_b$ changes much less. This result coincides quite well with that of Hall measurement, and it can be explained by the same mechanism mentioned in 3.1.

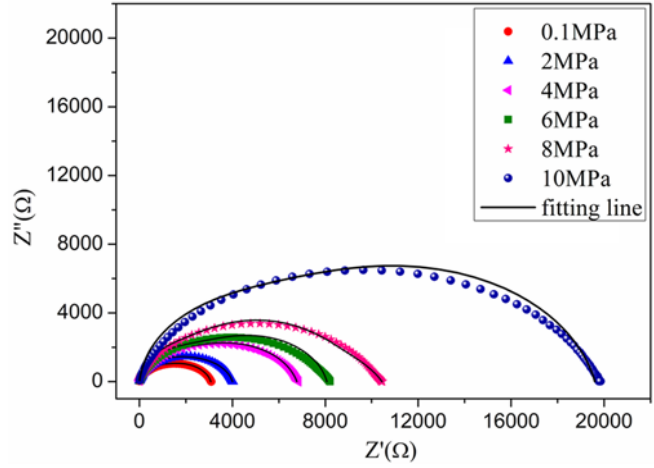

Fig. 9. Impedance spectra of samples calcined at different oxygen pressures

Table 4. Fitting results of impedance spectra of samples calcined at 500 °C for 2 h in high-pressure oxygen

| $SnO_2$ PNSs | $O_2$ pressure (MPa) | $R_e$ (Ω) | $R_{gb}$ (Ω) | $R_b$ (Ω) |
|---|---|---|---|---|
| P1 | 0.1 | 6.79 | 2882 | 1102 |
| P2 | 2.0 | 7.90 | 3442 | 1737 |
| P3 | 4.0 | 4.59 | 5871 | 2034 |
| P4 | 6.0 | 5.47 | 6770 | 2186 |
| P5 | 8.0 | 7.94 | 7398 | 5839 |
| P6 | 10.0 | 15.20 | 19170 | 6167 |

3.3 Further increasing the mobility of $SnO_2$ PNS via low-temperature long-duration calcination

Just as mentioned above, the carrier concentration, hence the conductivity of $SnO_2$ closely relates to the oxygen vacancies. Furthermore, it is reported that the annihilation of oxygen vacancies in $SnO_2$ happens at ~350 ºC, while the formation of oxygen vacancies appears at much higher temperature [24, 32]. In order to facilitate the annihilation of oxygen vacancies while suppressing their re-generation within $SnO_2$ nanoparticles, we calcined $SnO_2$ PNS at 350 ºC in high-pressure oxygen for 12 h, thus samples L1, L2, L3, L4, L5, L6 were obtained. The resistivity, carrier concentration and mobility of these samples were evaluated by Hall measurement data, and the results are presented in Table 5 and Fig. 10. Obviously, the mobility exhibits a maximum of 35 cm$^2$/Vs at 4 MPa, which is much higher than that of the samples calcined at 500 °C. On comparison, the carrier concentration decreases monotonically while increasing the oxygen pressure.

Table 5. Carrier concentration and Hall mobility of samples calcined at 350 ˚C for 12 h in high-pressure $O_2$

| $SnO_2$ PNSs | $O_2$ pressure (MPa) | Resistivity (Ω·cm) | $n \times 10^{-17}$ (1/cm³) | $\mu_H$ (cm²/Vs) |
|---|---|---|---|---|
| L1 | 0.1 | 0.73 | 4.82 | 17.66 |
| L2 | 2 | 0.71 | 3.77 | 23.21 |
| L3 | 4 | 1.04 | 1.72 | 35.00 |
| L4 | 6 | 2.77 | 1.20 | 18.82 |
| L5 | 8 | 8.37 | 1.29 | 5.79 |
| L6 | 10 | 12.86 | 1.04 | 4.88 |

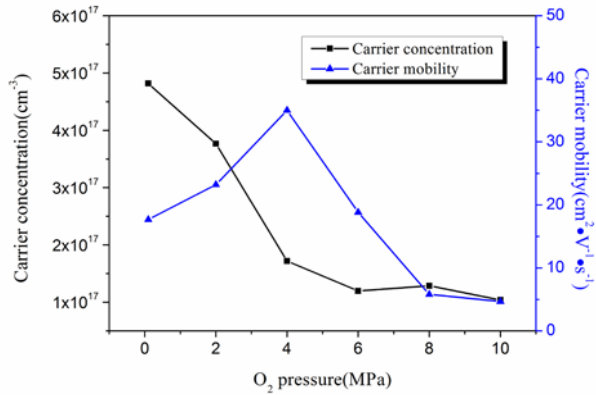

Fig. 10. Carrier concentration and mobility of $SnO_2$ PNSs calcined at 350 ˚C for 12 h in high-pressure $O_2$

It is well known that oxygen vacancies are the major scattering centers for the charge carriers in $SnO_2$. When being heated at 350 ˚C in high-pressure oxygen, three processes may have happened within $SnO_2$ PNS. The first one comprises the diffusion of oxygen vacancies towards the inner part of $SnO_2$ nanoparticles, thus the distribution of oxygen vacancies becomes more uniform, which decreases the interfacial potential barrier among $SnO_2$ nanoparticles. The second process includes the annihilation of a certain amount of oxygen vacancies, resulting in the improvement of crystallinity of $SnO_2$ nanoparticles, especially within the interfacial region [23, 24]. This phenomenon will decrease the interfacial potential barrier and increase the carrier mobility. The third process is that more and more oxygen molecules are chemisorbed on the surface of $SnO_2$ nanoparticles when the oxygen pressure gradually increasing, grabbing more electrons from $SnO_2$ nanoparticles and resulting in the thickening of depletion layer. As a result, the migration path of charge carriers becomes narrower [33], and the mobility decreases accordingly (Fig. 11). All these processes compete with each other, and reach equilibrium at a specific oxygen pressure, for example, 4 MPa in our sample.

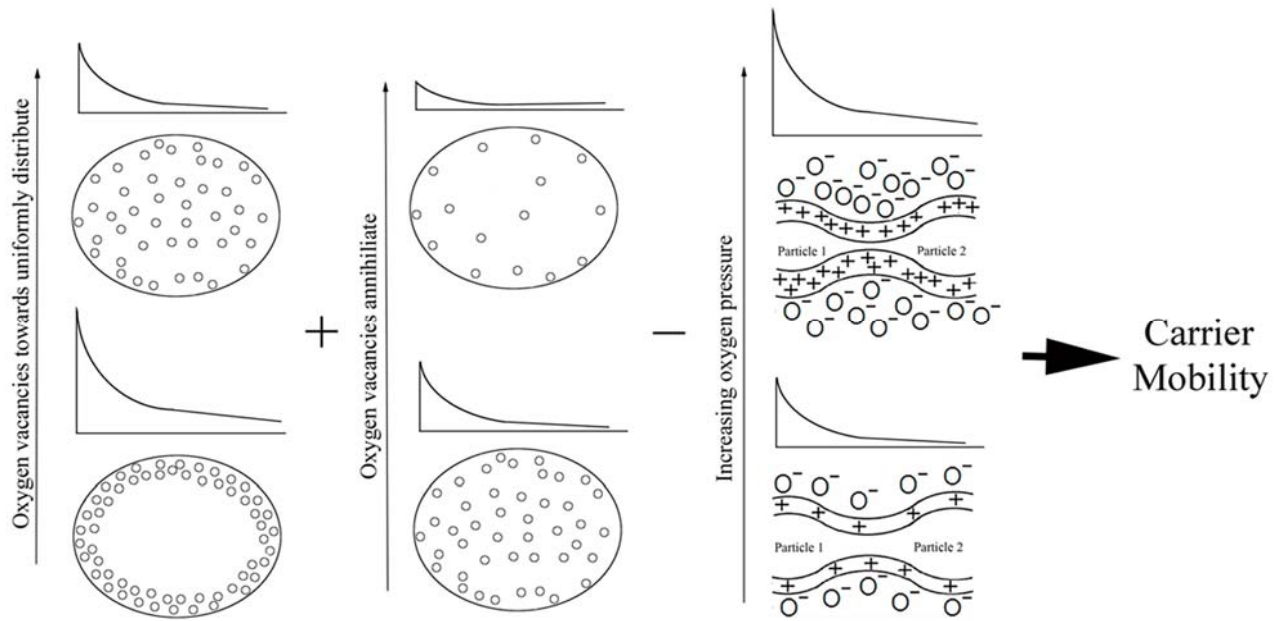

Fig. 11. A schematic diagram of the processes happened in $SnO_2$ PNS when being calcined in high-pressure oxygen

## 4 Conclusions

$SnO_2$ porous nanosolid was prepared using a solvothermal hot-press route, and its carrier mobility was modulated by calcining in different atmospheres. Especially, the mobility of a $SnO_2$ porous nanosolid greatly increased after being calcined in high-pressure oxygen. Furthermore, by thermal-treating $SnO_2$ porous nanosolid at 350 °C for 12 h in oxygen of 4 MPa, the mobility greatly increased to 35 $cm^2$/Vs. This phenomenon may find applications in improving the performance of photovoltaic cells based on $TiO_2$ ($SnO_2$)/organic dye composite materials.

**Acknowledgements**

This work is supported by the Natural Science Foundation of China (NSFC 50990061, 51021062, 21073107, 51102151), Natural Science Foundation (ZR2011EMQ002) and Postdoctoral Innovation Foundation (201003077) of Shandong Province, Independent Innovation Foundation (2010TS039) and Postdoctoral Foundation of Shandong University.